\documentclass{elsart}
\usepackage{graphicx,amssymb}

\usepackage{amsmath}

\begin{document}

\begin{frontmatter}



\title{Electron-Phonon interaction and electronic decoherence in molecular
conductors.}
\author[label1]{Horacio M. Pastawski}
\author[label1]{L. E. F. Foa Torres} and
\author[label2]{Ernesto Medina}

\address[label1]{Facultad de Matem\'{a}tica Astronom\'{\i}a y F\'{\i}sica,
Universidad Nacional de C\'{o}rdoba,Ciudad Universitaria, 5000 C\'{o}rdoba, Argentina.}
\address[label2]{Centro de F\'{\i}sica, Instituto
Venezolano de Investigaciones Cient\'{\i}ficas, Apartado 21827, 
Caracas 1020A, Venezuela.}

\begin{abstract}
We perform a brief but critical review of the Landauer picture of transport
that clarifies how decoherence appears in this approach. On this basis, we
present different models that allow the study of the coherent and decoherent
effects of the interaction with the environment in the electronic transport.
These models are particularly well suited for the analysis of transport in
molecular wires. The effects of decoherence are described through the
D'Amato-Pastawski model that is explained in detail. We also consider the
formation of polarons in some models for the electron-vibrational
interaction. Our quantum coherent framework allows us to study many-body
interference effects. Particular emphasis is given to the occurrence of
anti-resonances as a result of these interferences. By studying the phase
fluctuations in these soluble models we are able to identify inelastic and
decoherence effects. A brief description of a general formulation for the
consideration of time-dependent transport is also presented.
\end{abstract}

\end{frontmatter}


\section{Introduction.}

Electronic transport in biological and organic molecules has become a very
exciting field \cite{cit-molec-elect-NATURE,cit-RATNER-physics world} that
brings together ideas and results developed during the past two decades in
many branches of Physics, Chemistry and Biology. On the basis of this
synergic interaction one can foresee very innovative results. Much of its
wealth comes from the fact that molecules are intrinsically quantum objects,
and quantum mechanics always defies our classical intuition. When this
happens, we are almost certain to find new ``unexpected phenomena'' leading
to prospective applications. In turn, with a few exceptions, it is not yet
clear how Nature exploits quantum effects in biological systems. However, it
seems likely that evolution has made use of the details of the quantum
tunneling process in modulating charge transport \cite
{cit-RMP-frauenfelder-wolynes}. Hence, by pursuing an exploration of
dynamical quantum phenomena at the molecular level one may also expect to be
more prepared to discover Her hidden ways.

A paradigmatic example of the above situation is the electron transfer in
DNA strands \cite{cit-RATNER-physics world}. There is evidence that at least
two mechanisms are present in this case: 1) A coherent tunneling between the
base pairs constituting the donor and acceptor centers. 2) An inelastic
sequential hopping through bridging centers. Theoretically, there is a need
to unify the description of these extreme regimes as well as to study the
possible role of vibrations and distortions of the DNA structure. Our
present understanding of these basic physical processes comes from the field
of Quantum Transport, which evolved from the description of non-crystalline
materials \cite{cit-RMP-anderson} and reached its climax with nanoelectronic 
\cite{cit-Esaki} devices. Since these last structures may have a very small
length scale the term ``artificial atoms'' \cite{cit-Kastner} is amply
justified. Many of the phenomena appearing in these systems could be
summarized\ in the fact that the wave nature of electrons made them
susceptible to interference phenomena. These can be assimilated to the
propagation of light in some complex Fabry-Perrot interferometer producing
either well defined fringes or ``speckle-patterns''. Failures in that
simplistic description is essentially due to interactions with excessively
complex environmental degrees of freedom (such as thermal vibrations), and
our lack of control over them is interpreted as ``decoherence''\cite
{cit-Webb-decoh}. This often justifies the use of a classical description of
the transport process. However, if one is to borrow some idea from the
theory of solid state physics, it could be that coherent interactions
between electrons and lattice distortions give rise to new effects such as
assisted tunneling \cite{cit-T+e-ph/expt} or even superconductivity. Indeed
both phenomena could give striking results in organic systems \cite
{cit-assisted-tunn-MOL-exp,cit-Battlog}. One is then compelled to develop
new tools to describe electron-phonon (e-ph) coherent processes in molecular
devices.

In what follows we will make a brief description of quantum interference
phenomena establishing a ground level language based on the simplest
physical models. While following our personal pathway of many years through
the general ideas of transport in the quantum regime, we expect to induce a
new perspective into the subtle mechanisms that lead to the degradation of
the simple quantum effects through the interactions with the environment
(i.e. decoherence). As a token, we will visualize coherent effects emerging
from the electron-lattice interaction that can be exploited in new useful
ways.

In Section 1 we recall the Landauer ideas \cite{cit-Landauer} for transport
which will also serve to adopt a basic language and give a conceptual
framework into which the phenomenological aspects of decoherence can fit. In
Section 2 we review the D'Amato-Pastawski (DP) model \cite
{cit-D'Amato-Pastawski} for coherent and decoherent transport commenting its
strengths and limitations. In Section 3 we introduce a polaronic model that
hints at the properties of the complete electron-phonon Fock space. This not
only sheds light on the decoherence process but can also be used to predict
new phenomena such as the coherent emission of phonons. We devote Section 4
to a brief description of the time dependent transport. Section 5 gives a
general perspective.

\section{Landauer's picture for transport.}

\begin{figure}
\begin{center}
\includegraphics*[width=8cm]{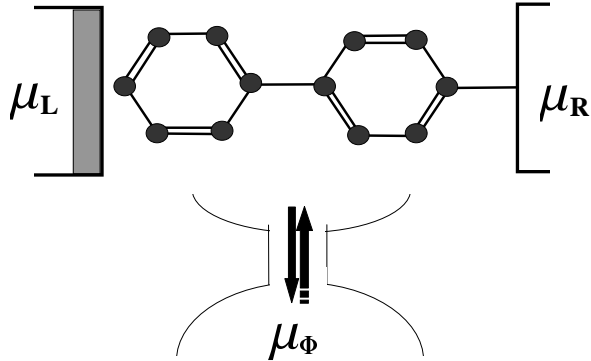}
\end{center}
\caption{Representation of a three probe measurement. The voltmeter may be strongly coupled and is a source of decoherence.}
\label{fig-3reservorios}
\end{figure}

The framework that inspired most of the described experimental developments
in mesoscopic electronics was the Landauer's \cite{cit-Landauer} simple but
conceptually new approach. Besides the ``sample'' or device, he explicitly
incorporated the description of the electric reservoirs connected to ``the
sample''. The role of reservoirs can be\ played not only by electrodes but
also could be spatial regions (localized LCAO) where the electrons lose
their quantum coherence. This last situation describes transport in the
hopping regime \cite{cit-Medina-MIT} (we will see more on this latter). The
simplest\ mathematical description is obtained if one thinks them as
one-dimensional wires that connect the individual orbitals $i$ to electronic
reservoirs characterized by the statistical distribution function $\mathrm{f}%
_{i}(\varepsilon )$. In that case, the electronic states are plane waves
describing the different boundary conditions of electrons ``in'' or ``out''
the reservoir. An electron ``out'' from reservoir connected to ``site'' $\ i$
has a probability $T_{j,i}(\varepsilon )$ to enter the reservoir connected
to ``site'' $j$.$\ $ A representation of such a situation for the case of
three reservoirs is sketched in Fig. \ref{fig-3reservorios}. The current per
spin state at reservoir $j$ is obtained by the application of the
Kirschhoff's law (i.e. a balance equation \cite{cit-Büttiker-Kirschhoff}): 
\begin{equation}
\mathsf{I}_{j}=\frac{e}{h}\int \mathrm{d}\varepsilon \sum_{i}\left[
T_{i,j}(\varepsilon )v_{j}\tfrac{1}{2}N_{j}\mathrm{f}_{j}(\varepsilon
)-T_{j,i}(\varepsilon )v_{i}\tfrac{1}{2}N_{i}\mathrm{f}_{i}(\varepsilon )%
\right] .  \label{eq-current-multilead}
\end{equation}
The meaning of this equation is obvious: it balances currents. Each
reservoir $j$ emits electrons with an energy availability controlled by a
Fermi distribution function $\mathrm{f}_{i}(\varepsilon )=1/[\exp
[(\varepsilon -\mu _{i})/k_{B}T+1]$, with a chemical potential, $\mu
_{i}=\mu _{o}+\delta \mu _{i}$, displaced from its equilibrium value $\mu
_{o}$. One can assimilate $\mathsf{V}_{i}=\delta \mu _{i}/e$ as voltages.
The density of those outgoing states is $\tfrac{1}{2}N_{i}(\varepsilon )$
(half the total) and their velocity $v_{i}$. It was essential in Landauer's
reasoning to note that in a propagating channel the density of states $N_{i}$
is inversely proportional to the corresponding group velocity: 
\begin{equation}
N_{i}\equiv 1/(v_{i}h).  \label{eq-N=1/veloc}
\end{equation}
This is immediately satisfied in one-dimensional wires but its validity is
much more general. This fundamental fact remained unnoticed in the early
discussions of quantum tunneling \cite{cit-Esaki} and it is the key to
understand conductance quantization. The $\mathsf{G}_{j,i}=\frac{e^{2}}{h}%
T_{j,i}(\varepsilon )$ are hence Landauer's conductances per spin channel.
For perfect transmitting samples $T_{j,i}$ is either 1 or 0 and one obtains
the conductance quantization in integer multiples of $e^{2}/h$.

Notice that there is no need for the traditional $\left[ 1-\mathrm{f}%
_{j}(\varepsilon )\right] $ factor to exclude transitions to already
occupied final states. In a scattering formulation, any ``in'' state
contains a linear combination of ``out'' states. Although two different
``in'' states (e.g. on the left and right electrodes) could end in the same
final ``out'' state, unitarity of quantum mechanics assures that both sets
are orthogonal. Here the transmission coefficients may depend on the
external parameters such as voltages.

An important particularity of our Eq. (\ref{eq-current-multilead}) is that
it does not exclude sites $i=j$ from the sum. This contrasts with the
original multichannel description\cite{cit-Büttiker-Kirschhoff} and is of
utmost importance in the treatment of time dependent problems where $\mathsf{%
V}_{i}=\mathsf{V}_{i}(t)$ as will be seen in Section 4. Our Eq. (\ref
{eq-current-multilead}), always used with Eq. (\ref{eq-N=1/veloc}), has a
full quantum foundation within the Keldysh formulation of Quantum Mechanics
(e.g. see Eqs. (5.6-7) in ref. \cite{cit-GLBE2}). It can be fully expressed
in terms of quantities obtained from a Hamiltonian model such as local
density of states and Green's functions. We adopt here a notation consistent
with these formal developments.

\subsection{Phenomenology of decoherence.}

\begin{figure}
\begin{center}
\includegraphics*[width=9cm]{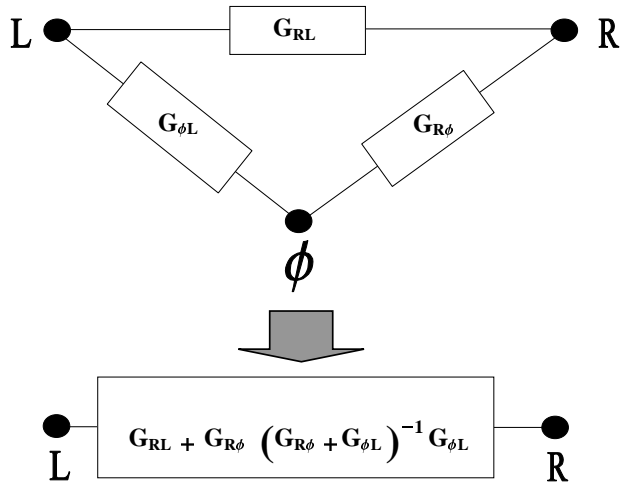}
\end{center}
\caption{Classical circuit representation of the non-classical system of Fig. 1.}
\label{fig-resistencias}
\end{figure}

A first alternative to include decoherence in steady state quantum transport
was inspired in the Landauer's formulation. There, the leads, while
accepting a quantum description of their spectrum of propagating
excitations, are the ultimate source of irreversibility and decoherence:
electrons leaving the electrodes toward the sample are completely incoherent
with the electrons coming from the other electrodes. In fact, it is obvious
that a wire connected to a voltmeter, by ``measuring'' the number of
electrons in it, must produce some form of \ ``collapse'' of the wave
function leading to decoherence (see Fig. \ref{fig-3reservorios}). Besides,
no net current flows toward a voltmeter. The leads are then a natural source
of decoherence which can be readily described in the Landauer picture if one
uses the Landauer conductances together with the Kirschhoff balance
equations. This fact was firstly realized by M. B\"{u}ttiker \cite
{cit-Büttiker-decoh}. Let us see how it works for the case of a single
voltmeter in the linear response regime. In matrix form: 
\begin{equation}
\left( 
\begin{array}{l}
\mathsf{I}_{\mathrm{L}} \\ 
\mathsf{I}_{\phi } \\ 
\mathsf{I}_{\mathrm{R}}
\end{array}
\right) =\left[ 
\begin{array}{lll}
-\left[ \mathsf{G}_{\mathrm{R,L}}+\mathsf{G}_{\phi ,\mathrm{L}}\right] & 
\,\,\,\ \,\,\,\,\,\,\,\,\ \,\mathsf{G}_{\mathrm{L},\phi } & \,\,\,\
\,\,\,\,\,\,\,\,\mathsf{G}_{\mathrm{L,R}} \\ 
\,\,\,\ \,\,\,\,\,\,\,\,\ \,\mathsf{G}_{\phi ,\mathrm{L}} & -\left[ \mathsf{G%
}_{\mathrm{R},\phi }+\mathsf{G}_{\mathrm{L},\phi }\right] & \,\,\,\
\,\,\,\,\,\,\,\,\,\mathsf{G}_{\phi ,\mathrm{R}} \\ 
\,\,\,\ \,\,\,\,\,\,\,\,\ \,\mathsf{G}_{\mathrm{R,L}} & \,\,\,\
\,\,\,\,\,\,\,\,\ \,\mathsf{G}_{\mathrm{R},\phi } & -\left[ \mathsf{G}_{\phi
,\mathrm{R}}+\mathsf{G}_{\mathrm{L,R}}\right]
\end{array}
\right] \left[ \left( 
\begin{array}{l}
\mathsf{V}_{\mathrm{L}} \\ 
\mathsf{V}_{\phi } \\ 
\mathsf{V}_{\mathrm{R}}
\end{array}
\right) \right] .
\end{equation}
Here the unknowns are $\mathsf{I}_{\mathrm{L}}$, $\mathsf{I}_{\mathrm{R}}$
and $\mathsf{V}_{\phi }=\delta \mu _{\phi }/e$. The second equation must be
solved with the voltmeter condition $\mathsf{I}_{\phi }\equiv 0$: 
\begin{equation}
0=\frac{e}{h}T_{\phi ,\mathrm{L}}(\delta \mu _{\phi }-\delta \mu _{\mathrm{L}%
})+\frac{e}{h}T_{\mathrm{R},\phi }(\delta \mu _{\phi }-\delta \mu _{\mathrm{R%
}}),
\end{equation}
giving us $\delta \mu _{\phi }$ to be introduced in the third equation:

\begin{equation}
\mathsf{I}_{\mathrm{R}}=\frac{e}{h}T_{\mathrm{R,L}}(\delta \mu _{\mathrm{L}%
}-\delta \mu _{\mathrm{R}})+\frac{e}{h}T_{\mathrm{R},\phi }(\delta \mu
_{\phi }-\delta \mu _{\mathrm{R}})
\end{equation}
to obtain the current 
\begin{equation*}
\mathsf{I}_{\mathrm{R}}=\frac{e}{h}\widetilde{T}_{\mathrm{R,L}}(\delta \mu _{%
\mathrm{L}}-\delta \mu _{\mathrm{R}})
\end{equation*}
with 
\begin{equation}
\widetilde{T}_{\mathrm{R},\mathrm{L}}=T_{\mathrm{R},\mathrm{L}}+\frac{T_{%
\mathrm{R},\phi }T_{\phi ,\mathrm{L}}}{T_{\mathrm{R},\phi }+T_{\phi ,\mathrm{%
L}}}  \label{eq-incoherent-transmittance}
\end{equation}
The first term can be identified with the coherent part, while the second is
the incoherent or sequential part, i.e. the contribution to the current
originated from particles that interact with the voltmeter. This corresponds
to an effective conductance of 
\begin{equation}
\widetilde{\mathsf{G}}_{\mathrm{R,L}}=\mathsf{G}_{\mathrm{R,L}}+(\mathsf{G}_{%
\mathrm{R},\phi }^{\,\,\,\,\,\,\,\,\,\,-1}+\mathsf{G}_{\phi ,\mathrm{L}%
}^{\,\,\,\,\,\,\,\,\,\,-1})^{-1},  \label{eq-paralel-resistances}
\end{equation}
which can be identified with the electrical circuit of Fig. \ref
{fig-resistencias}. This classical view clarifies the competition between
coherent and incoherent transport. However, this circuit does not imply that
in Quantum Mechanics one cannot modify one of the resistances without deeply
altering the others. This fact can become very relevant in partially
coherent regimes.

So far with the phenomenology. The next important step is to connect these
quantities with actual model Hamiltonians. This connection was made explicit
by the contribution of D'Amato and Pastawski \cite{cit-D'Amato-Pastawski}.

\section{The D'Amato-Pastawski model for decoherence.}

\subsection{Reducing the Hamiltonian}

Let us first review the basic mathematical background that made possible the
selection of a simple Hamiltonian that best represents the complex
sample-environment system. The objective was to find a simple way to account
for the infinite degrees of freedom of a thermal bath and/or electrodes and
use the exact solution in the Landauer's transport equation. First, we
recall that one can always eliminate the microscopic degrees of freedom \cite
{cit-Löwdin,cit-Levstein-decim} generating an effective Hamiltonian,
expressed in a basis of localized orbitals. This produces effective
interactions and energy renormalizations which depend themselves on the
observed energy. Furthermore, one can include a whole lead in a Hamiltonian
description through a correction to the eigen-energies which has an \textit{%
imaginary }part. In fact, an electron originally localized in the region
called ``the sample'' should eventually escape or decay toward the
electrodes. In a microscopic description this is equivalent to the decay of
an excited atom according to the Fermi Golden Rule. Hence, there is a escape
velocity associated with the energy uncertainty of a local state: 
\begin{equation}
v_{i}=\frac{2a}{\hbar }\Gamma _{i}=\dfrac{a}{\tau _{i}},  \label{eq-v-Gamma}
\end{equation}
where $\ a$ is a lattice constant. Let us see a simple example. Consider a
single LCAO which could be the highest occupied molecular orbital (HOMO) or 
the lowest unoccupied molecular orbital (LUMO) responsible for resonant
transport 
\begin{equation*}
\hat{\mathcal{H}}_{\mathrm{0}}^{o}=E_{0}\widehat{c}_{0}^{+}\widehat{c}%
_{0}^{{}}
\end{equation*}
coupled with a quasi-\textit{continuum} of electronic states of the
electrode. They are described by a ``lead'' connected to the voltmeter: 
\begin{equation*}
\hat{\mathcal{H}}_{\mathrm{lead}}=\sum_{k}E_{k}^{{}}\widehat{c}_{k}^{+}%
\widehat{c}_{k}^{{}}.
\end{equation*}
The sample-lead interaction term is: 
\begin{equation*}
\hat{\mathcal{H}}_{\mathrm{0-lead}}=\sum_{k}\left( V_{k,0}\widehat{c}_{k}^{+}%
\widehat{c}_{0}^{{}}+\mathrm{c.c.}\right) .
\end{equation*}
The unperturbed atomic energy $E_{0}$ will become corrected by the presence
of the lead. In second order of perturbation we get 
\begin{equation}
^{\mathrm{lead}}\Sigma _{0}^{R(A)}=\lim_{\eta \rightarrow 0^{+}}\sum_{k}%
\frac{\left| V_{k,o}\right| ^{2}}{E_{0}-E_{k}\pm \mathrm{i}\eta }=\,^{%
\mathrm{lead}}\Delta _{0}(E_{0})\mp \mathrm{i}^{\mathrm{lead}}\Gamma
_{0}(E_{0}).  \label{eq-Sigma}
\end{equation}
The sign $+$ or $-$ of the infinitesimal imaginary energy $\eta $ \ is
introduced to handle eventual divergencies in the sum. Since it determines
the sign of time in the evolution, the supra-index $R$ or $A$ corresponds to
either a retarded or advanced propagation. The imaginary component appears
because the electrode spectrum, described by the density of states $N_{%
\mathrm{lead}}(\varepsilon )$, is continuum in the neighborhood of energy $%
E_{o}+\Delta (E_{o}).$ This makes possible the irreversible \textit{decay}
into the continuum set of states $k$, not included in the bounded
description. The general functional dependence of $\Delta _{0}(E_{o})-%
\mathrm{i}\Gamma _{0}(E_{o})$ is better expressed in terms of the
Wigner-Brillouin perturbation theory: 
\begin{equation}
^{\mathrm{lead}}\Delta _{0}(\varepsilon )=\wp \int_{-\infty }^{\infty }\frac{%
\left| V_{k,0}\right| ^{2}}{\varepsilon -E_{k}}N_{\mathrm{lead}}(E_{k})%
\mathrm{d}E_{k},
\end{equation}
where $\wp $ stands for principal value and $N_{\mathrm{lead}}(E_{k})$ is
the density of states at the lead. Similarly: 
\begin{equation}
^{\mathrm{lead}}\Gamma _{0}(\varepsilon )=\pi \int_{-\infty }^{\infty
}\left| V_{k,0}\right| ^{2}N_{\mathrm{lead}}(E_{k})\delta \lbrack
\varepsilon -E_{k}]\mathrm{d}E_{k}.  \label{eq-gamma}
\end{equation}
The evaluation of Eq. (\ref{eq-gamma}) and (\ref{eq-v-Gamma}) at $%
\varepsilon =E_{0}$ constitutes the Fermi Golden Rule (FGR). Of course these
quantities satisfy the Kramers-Kroning relations

\begin{equation}
\Delta _{0}(\varepsilon )=\frac{1}{\pi }\wp \int_{-\infty }^{\infty }\frac{%
\Gamma _{0}(\varepsilon ^{\prime })}{\varepsilon -\varepsilon ^{\prime }}%
\mathrm{d}\varepsilon ^{\prime }.
\end{equation}
An explicit functional dependence on the variable $\varepsilon $ contains
eventual non-FGR behavior. The FGR describes reasonably electrons in an atom
decaying into the continuum (electrode states), or propagating electrons
decaying into different momentum states by collision with impurities or
interaction with a field of phonons or photons. In some of these cases, we
have to add some degrees of freedom to the sum (the phonon or photon
coordinates). A process $\alpha $ may produce contributions $^{\alpha
}\Sigma _{0}^{R}$ to the total self-energy $\Sigma _{0}^{R}$. 
\begin{equation}
\widehat{\mathcal{H}}_{0}^{\circ }{}_{\overrightarrow{\mathrm{interactions}}}%
\widehat{\mathcal{H}}_{0}=\widehat{\mathcal{H}}_{0}^{(\circ )}+\widehat{%
\Sigma }_{0}^{R}(\varepsilon )
\end{equation}
with

\begin{eqnarray}
\widehat{\Sigma }_{0}^{R(A)}(\varepsilon ) &=&\left( \Delta _{0}(\varepsilon
)\mp \mathrm{i}\Gamma _{0}(\varepsilon )\right) \widehat{c}_{0}^{+}\widehat{c%
}_{0}^{{}}  \label{eq-sigma-equal-to-delta-igama} \\
&=&\sum_{\alpha }\left( ^{\alpha }\Delta _{0}(\varepsilon )\mp \mathrm{i\,\,}%
^{\alpha }\Gamma _{0}(\varepsilon )\right) \widehat{c}_{0}^{+}\widehat{c}%
_{0}^{{}}.  \label{eq-Sigma-suma-procesos}
\end{eqnarray}
Besides, the best way to do perturbation theory to infinite order is the
framework of Green's functions.

\begin{figure}
\begin{center}
\includegraphics*[width=10cm]{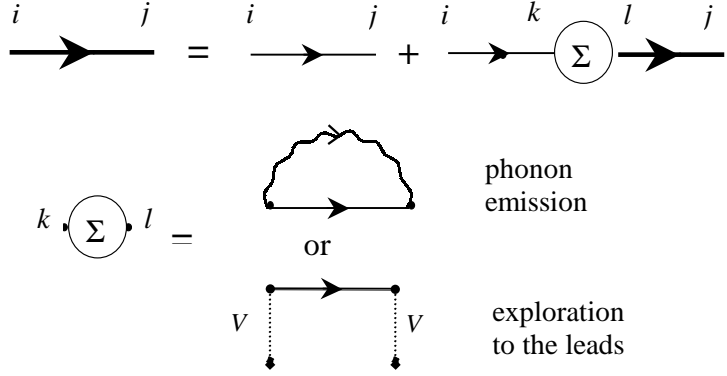}
\end{center}
\caption{Feynman representation of the Dyson equation for the single particle Green's function (line). An electron-phonon self-energy is evaluated in terms of the sample electron (line) and phonon (wave) Green's functions. The leads self-energies contain the hopping (dot) and the propagator in the lead. }
\label{fig-feynmanG}
\end{figure}

The unperturbed retarded Green's functions are defined as the matrix
elements of the resolvent operator 
\begin{equation}
\widehat{G}^{oR}(\varepsilon )=\lim_{\eta \rightarrow 0^{+}}[(\varepsilon +%
\mathrm{i}\eta )\hat{I}-\widehat{\mathcal{H}}_{{}}^{o}]^{-1}
\end{equation}
for the \textit{isolated} sample. It is practical to represent this as a
matrix, $\mathbf{G}^{oR}(\varepsilon )$, whose elements are written in terms
of the eigen-energies $E_{\alpha }^{o}$ and eigen-functions $\psi _{\alpha
}^{(o)}\left( \mathbf{r}\right) =\sum_{i}u_{\alpha ,i}\varphi _{i}(\mathbf{r}%
)$ of the isolated sample as 
\begin{equation}
G_{i,j}^{oR}(\varepsilon )=\lim_{\eta \rightarrow 0^{+}}\sum_{\alpha }\frac{%
u_{\alpha ,i}^{{}}u_{\alpha ,j}^{\ast }}{\varepsilon +\mathrm{i}\eta
-E_{\alpha }^{o}}=\left[ G_{j,i}^{oA}(\varepsilon )\right] ^{\ast }
\label{eq-G-eigenfunctions}
\end{equation}
$\,$The Local Density of States at orbital $i$ is calculated as 
\begin{equation}
N_{i}^{o}=\dfrac{1}{2\pi }\lim_{\eta \rightarrow
0^{+}}[G_{i,i}^{oA}(\varepsilon )-G_{i,i}^{oR}(\varepsilon )].
\label{eq-local-densities}
\end{equation}
The Fourier transform 
\begin{equation}
G_{i,j}^{oR}(t_{2}-t_{1})=\int_{-\infty }^{\infty }\frac{\mathrm{d}%
\varepsilon }{2\pi \hbar }\exp \left[ -\frac{\mathrm{i}}{\hbar }\varepsilon
(t_{2}-t_{1})\right] G_{i,j}^{oR}(\varepsilon )
\end{equation}
is solution of: 
\begin{equation}
\left[ \left( -\mathrm{i}\hbar \dfrac{\partial }{\partial t_{2}}\mathbf{I}+%
\mathbf{H}^{o}\right) \mathbf{G}_{{}}^{oR}(t_{2}-t_{1})\right] _{i,j}=\delta
_{i,j}\delta (t_{2}-t_{1})  \label{eq-retarded-Greens}
\end{equation}
with $t_{2}>t_{1}.$ The identity matrix is $\left[ \mathbf{I}\right]
_{i,j}=\delta _{i,j}$. Hence $G_{i,j}^{oR}(t_{2}-t_{1})$ is $-\tfrac{\mathrm{%
i}}{\hbar }$ times the probability amplitude that a particle placed in $j$%
-th orbital at time $t_{1}$ be found in $i$-th one at time $t_{2}$. The
advantage of the method is that $G_{i,j}^{R}(t)$ can be calculated
numerically from $G_{i,j}^{R}(\varepsilon )$ without a detailed knowledge of
the eigen-solutions of the perturbed system. For $G_{i,j}^{R}(\varepsilon )$
one uses: 
\begin{align}
\mathbf{G}^{R}(\varepsilon )& =[\varepsilon \mathbf{1}-(\mathbf{H}_{o}+%
\mathbf{\Sigma }^{R}(\varepsilon ))]^{-1}  \label{eq-G-self-energy} \\
& =\mathbf{G}^{oR}(\varepsilon )+\mathbf{G}^{oR}(\varepsilon
)\sum_{n=1}^{\infty }\left[ \mathbf{\Sigma }^{R}(\varepsilon )\mathbf{G}%
^{oR}(\varepsilon )\right] ^{n} \\
& =\mathbf{G}^{oR}(\varepsilon )+\mathbf{G}^{oR}(\varepsilon )\,\mathbf{%
\Sigma }^{R}\,(\varepsilon )\mathbf{G}^{R}(\varepsilon )
\label{eq-Dyson-G-Sigma}
\end{align}
In the second line we have written the usual forms of the Dyson equation
from which the Wigner-Brillouin perturbative series and its representation
in Feynman diagrams can be obtained. In Fig. \ref{fig-feynmanG} we show the
graphical representation of the Dyson equation considering two local
contributions to the self-energy: the electron-phonon interactions and the
escape to the leads.

For a brief tutorial on the calculation of the Green's function in discrete
systems see Ref. \cite{cit-horacio-REVIEW-electransp}. Although the
formalism seems to introduce some extra notation it has various conceptual
advantages. For example, it is straightforward to use Eq. (\ref
{eq-G-self-energy}) to prove the \textit{optical theorem }\cite
{cit-horacio-REVIEW-electransp}: 
\begin{equation}
\lbrack \mathbf{G}^{R}-\mathbf{G}^{A}]=\mathbf{G}^{R}\,[\mathbf{\Sigma }^{R}-%
\mathbf{\Sigma }^{A}]\,\mathbf{G}^{A}  \label{eq-optical-theorem}
\end{equation}
of deep physical significance since it is an integral equation relating the
local densities of states given by Eq. (\ref{eq-local-densities}) and the
decay rates provided by Eq. (\ref{eq-sigma-equal-to-delta-igama}). Perhaps
the most important advantage, is that they can be used also in the Quantum
Field Theory \cite{cit-Keldysh,cit-Danielewicz} to deal with the many-body
case.

The direct connection between observables and the transmittances used in the
Landauer formulation and the Green's function follows intuitively from the
Green's function probabilistic interpretation. This was formalized by
D'Amato and Pastawski who used a result of Fisher and Lee \cite
{cit-Fisher-Lee} that related the transmittance to a product of the Green's
function connecting two regions and their group velocities. By noting the
discussed relation between the group velocity and the imaginary part of the
effective potential established in Eq. (\ref{eq-v-Gamma}), DP obtained a
simple expression for the transmittance, which in our present notation can
be written as 
\begin{equation}
T_{j\alpha ,i\beta }(\varepsilon )\equiv \left| t_{j\alpha ,i\beta
}(\varepsilon )\right| ^{2}=\left[ 2\,^{\alpha }\Gamma _{j}(\varepsilon )%
\right] \,G_{j,i}^{R}(\varepsilon )\,\left[ 2\,^{\beta }\Gamma
_{i}(\varepsilon )\right] G_{i,j}^{A}(\varepsilon )]\,\,.
\label{eq-Fisher&Lee}
\end{equation}
The left supra-index $\alpha $ in $^{\alpha }\Gamma $ indicates each of the 
\textit{independent} processes producing the decay from the LCAO modes
(right subindex) associated with the physical channel. In the basis of
independent channels the complex part of the self-energy is diagonal. Hence
the sum over initial states and processes in the physical channels $\mathrm{R%
}$ and $\mathrm{L}$ of Eq. (\ref{eq-current-multilead}) determines a
conductance: 
\begin{equation}
\mathsf{G}_{\mathrm{R,L}}=\dfrac{2e^{2}}{h}4\mathrm{Tr}[\mathbf{\Gamma }_{%
\mathrm{R}}(\varepsilon )\,\mathbf{G}_{\mathrm{R,L}}^{R}(\varepsilon )\,%
\mathbf{\Gamma }_{\mathrm{L}}(\varepsilon )\mathbf{G}_{\mathrm{L,R}%
}^{A}(\varepsilon )].
\end{equation}
This matrix expression is simply a compact way to write Eq.(\ref
{eq-current-multilead}), using Eqs. (\ref{eq-Fisher&Lee}) and (\ref
{eq-v-Gamma}) together to compute the linear response conductance between
any pair $\mathrm{L}$ and $\mathrm{R}$ of electrodes. The sum of initial
states at left ($\mathrm{L}$) is the result of the product of the diagonal $%
\mathbf{\Gamma }_{\mathrm{L}}(\varepsilon )$ form of the broadening matrix,
while the final trace is the sum over final states at right ($\mathrm{R}$).
With different notations, these basic ideas recognize by now many
applications to the field of molecular electron transfer \cite
{cit-Ratner-and-ETgroup}.

Originally, Fisher and Lee considered only the escape velocity to the leads
(i.e. $\alpha =\beta =\mathrm{lead}$). Our point is that \textit{any other
process} which contributes to the decay \textit{giving an imaginary}
contribution to the \textit{self-energy} would be described by Eq. (\ref
{eq-Fisher&Lee}). In particular, this will be true for a ``decoherence''
velocity that degrades the coherent current. This view was proposed in DP 
\cite{cit-D'Amato-Pastawski} adopting a discrete (tight-binding) description
of the spatial variables at each point (or orbital) in the real space a
decay rate was assigned which is balanced by the particles reinjection in
the same site, described as local reservoirs.

We now show how the DP model applies to a simple tunneling system. Consider
the unperturbed Hamiltonian, 
\begin{equation}
\hat{\mathcal{H}}_{0}=\sum_{i=0}^{N+1}\left\{ E_{i}\widehat{c}_{i}^{+}%
\widehat{c}_{i}^{{}}+\sum_{j\left( \neq i\right) }^{{}}V_{i,j}\left[ 
\widehat{c}_{i}^{+}\widehat{c}_{j}^{{}}+\widehat{c}_{j}^{+}\widehat{c}%
_{i}^{{}}\right] \right\} .
\end{equation}
Since the indices $i$ and $j$ (sites) refer to any set of atomic orbitals,
the interactions are not restricted to nearest neighbors. However, for the
usual short range interactions, the Hamiltonian matrix has the advantage of
being sparse. The local dephasing field is represented by 
\begin{equation}
^{\phi }\hat{\Sigma}^{R}=\sum_{i=1}^{N}-\mathrm{i\,\,}^{\phi }\Gamma \,\,%
\widehat{c}_{i}^{+}\widehat{c}_{i}^{{}}  \label{eq-DP model}
\end{equation}
where $^{\phi }\Gamma =\hbar /(2\tau _{\phi }).$ We consider for simplicity
only two one-dimensional current leads $\mathrm{L}$ and $\mathrm{R}$
connected to the $1$st. and the $N$th orbital states respectively, 
\begin{equation}
^{leads}\hat{\Sigma}^{R}=-\mathrm{i}\left( ^{\mathrm{L}}\Gamma \,\,\widehat{c%
}_{1}^{+}\widehat{c}_{1}+^{\mathrm{R}}\Gamma \,\,\widehat{c}_{N}^{+}\widehat{%
c}_{N}^{{}}\right) .
\end{equation}
We see that the $1$-st state has escape contributions both, toward the
current lead at the left, $^{\mathrm{L}}\Gamma _{1}$, and to the inelastic
channel associated to this site, $^{\phi }\Gamma _{1}$. The on-site chemical
potential will ensure that no net current flows through this channel.

\subsection{The solution for incoherent transport}

To simplify the notation we define the total transmission from each site as: 
\begin{equation}
\left( 1/g_{i}\right) =\sum_{j=0}^{N+1}T_{j,i}=\left\{ 
\begin{array}{c}
4\pi N_{1}{}^{\mathrm{L}}\Gamma _{1}\,\,\,\,\,\,\,\,\,\,\,\,\,\,\,\,\,\,%
\mathrm{for\,}i=0 \\ 
\,\,4\pi N_{i}{}^{\mathrm{\phi }}\Gamma _{1}\,\,\,\,\,\,\,\mathrm{for\,}%
1\leq i\leq N \\ 
\,\,4\pi N_{N}{}^{\mathrm{R}}\Gamma _{N}\,\,\,\,\,\mathrm{for\,}i=N+1
\end{array}
\right.  \label{eq-total-escape}
\end{equation}

The last equality follows from the optical theorem of Eq.(\ref
{eq-optical-theorem}). The balance equation becomes 
\begin{equation}
\mathsf{I}_{i}\equiv 0=-\left( 1/g_{i}\right) \delta \mu
_{i}+\sum_{j=0}^{N+1}T_{i,j}\delta \mu _{j},  \label{eq-localcurrents}
\end{equation}
where the sum adds all the electrons that emerge from a last collision at
other sites ($j$ 's) and propagate coherently to site $i$ where they suffer
a dephasing collision. These include the electrons coming from the current
source i.e. $T_{i,\mathrm{L}}\delta \mu _{\mathrm{L}}$ and the current
drain. However, since we refer all voltages to the last one, $T_{i,\mathrm{R}%
}\delta \mu _{\mathrm{R}}\equiv 0$. The first term accounts for all the
electrons that emerge from this collision on site $i$ to have a further
dephasing collision either in the sample or in the leads. The net current is
identically zero at any dephasing channel (lead). The other two equations
are 
\begin{align}
\mathsf{I}_{\mathrm{L}}& \equiv -\mathsf{I}=-\left( 1/g_{\mathrm{L}}\right)
\delta \mu _{\mathrm{L}}+\sum_{j=0}^{N}T_{\mathrm{L},j}\delta \mu _{j},
\label{eq-currentleads} \\
\mathsf{I}_{\mathrm{R}}& \equiv \mathsf{I}=-\left( 1/g_{\mathrm{R}}\right)
\delta \mu _{\mathrm{R}}+\sum_{j=0}^{N+1}T_{\mathrm{R},j}\delta \mu _{j}. 
\notag
\end{align}
Here we need the local chemical potentials which can be obtained from Eq. (%
\ref{eq-localcurrents}). In a compact notation, these coefficients can be
arranged in a matrix form which excludes the leads that are current source
and sink: 
\begin{equation}
\mathbf{W}=\left[ 
\begin{array}{lllll}
1/g_{1}-T_{1,1} & T_{1,2} & T_{1,3} & \cdots & T_{1,N} \\ 
T_{2,1} & 1/g_{2}-T_{2,2} & T_{2,3} & \cdots & T_{2,N} \\ 
T_{3,1} & T_{3,2} & 1/g_{3}-T_{3,3} & \cdots & T_{3,N} \\ 
\vdots & \vdots & \vdots &  & \vdots \\ 
T_{N,1} & T_{N,2} & T_{N,3} & \cdots & 1/g_{N}-T_{N,N}
\end{array}
\right] .
\end{equation}
Then, the chemical potential in each site can be calculated as 
\begin{equation}
\delta \mu _{i}=\sum_{j=1}^{N}\left( \mathbf{W}^{-1}\right)
_{i,j}T_{j,0}\delta \mu _{0}.
\end{equation}
Replacing these chemical potentials back in Eq. (\ref{eq-localcurrents}) the
effective transmission can be calculated 
\begin{equation}
\widetilde{T}_{\mathrm{R},\mathrm{L}}=T_{\mathrm{R},\mathrm{L}%
}+\sum_{j=1}^{N}\sum_{i=1}^{N}T_{\mathrm{R},j}\left[ \mathbf{W}^{-1}\right]
_{j,i}T_{i,\mathrm{L}}.  \label{eq-total transmision}
\end{equation}

The first contribution in the RHS comes from electrons that propagate
quantum \textit{coherently} through the sample. The second term contains the 
\textit{incoherent} contributions due to electrons that suffer their first
dephasing collision at site $i$ and their \textit{last one} at site $j$.

\begin{figure}
\begin{center}
\includegraphics*[width=12cm]{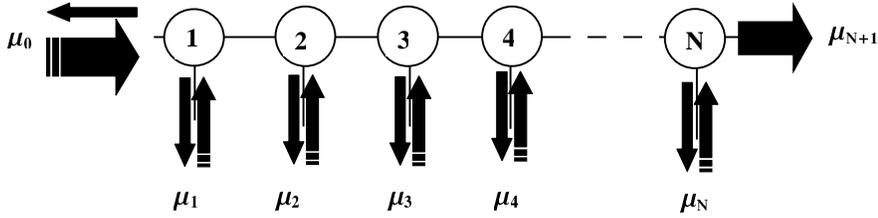}
\end{center}
\caption{Pictorial representation of the D Amato-Pastawski
model for the case of a linear chain.}
\label{fig-DAmato}
\end{figure}

Until now the procedure has been completely general, there is no assumption
involving the dimensionality or geometry of the sample. The system of Fig. 
\ref{fig-DAmato} was adopted in DP only because it has a simple analytical
solution for $T_{i,j}$ in various situations ranging from tunneling to
ballistic transport. We summarize the procedure for the linear response
calculation. First, we calculate the complete Green's function in a
tight-binding model. Since the Hamiltonian is sparse this is relatively
inexpensive if one uses the decimation techniques described in Ref. \cite
{cit-Levstein-decim}. With the Green's functions, we evaluate the
transmittances between every pair of sites in the sample (i.e. nodes in the
discrete equation) and write the transmittance matrix \textbf{W}. Then, we
solve for the current conservation equations that involve the inversion of 
\textbf{W}.

\begin{figure}
\begin{center}
\includegraphics*[width=10cm]{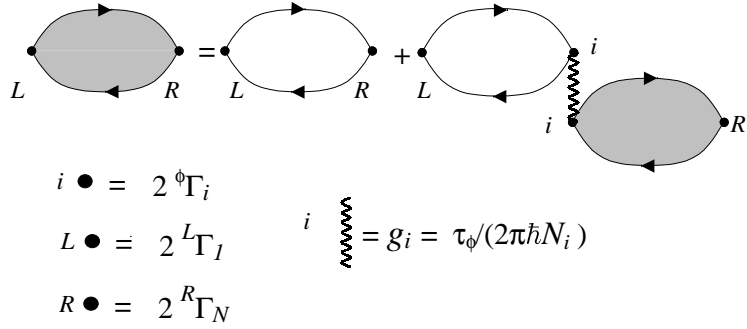}
\end{center}
\caption{Feynman Diagram for the Dyson Equation of the transmittance. It is equivalent to a particle-hole Green's function in the ladder approximation where a rung is represented by a dot.}
\label{fig-feynmanT}
\end{figure}

What are the limitations of this model? A conceptual one is the momentum
demolition produced by the localized scattering model. Therefore, by
decreasing $\tau _{\phi },$ the dynamics is transformed continuously from
quantum ballistic to classical diffusive. To describe the transition from
quantum ballistic to classical ballistic, one should modify the model to
have the scattering defined in phase space or energy basis. While the first
is well suited for scattering matrix models \cite{cit-k-conserving}, the
last is quite straightforward as will be shown in next Section. The other
aspect is merely computational. Since the resulting matrix $\mathbf{W}$ is
no longer sparse, this inversion is done at the full computational cost. A
physically appealing alternative to matrix inversion was proposed in DP. The
idea was to expand the inverse matrix in series in the dephasing collisions,
resulting in: 
\begin{align}
\widetilde{T}_{\mathrm{R},\mathrm{L}}& =T_{\mathrm{R},\mathrm{L}}+\sum_{i}T_{%
\mathrm{R},i}g_{i}T_{i,\mathrm{L}}+\sum_{i}\sum_{j}T_{\mathrm{R}%
,i}g_{i}T_{i,j}g_{j}T_{i,\mathrm{L}}  \label{eq-T-series} \\
& +\sum_{i}\sum_{j}\sum_{l}T_{\mathrm{R},i}g_{i}T_{i,j}g_{j}T_{i,l}g_{l}T_{l,%
\mathrm{L}}+\ldots  \notag
\end{align}
The formal equivalence with the self-energy expansion in terms of locators
or local Green's function justifies identifying $g_{i}$ as a \textit{locator}
for the classical Markovian equation for a density excitation \cite
{cit-GLBE1} generated by the transition probabilities $T^{\prime }$s. Notice
that Eq. (\ref{eq-T-series}) can also be rearranged as:

\begin{equation}
\widetilde{T}_{\mathrm{R},\mathrm{L}}=T_{\mathrm{R},\mathrm{L}%
}+\sum_{i=1}^{N}\widetilde{T}_{\mathrm{R},i}g_{i}T_{i,\mathrm{L}}.
\end{equation}
This has the structure of the Dyson equation, graphically represented in
Fig. \ref{fig-feynmanT}. We notice that according to the optical theorem $%
g_{i}=\tau _{\phi }2\pi \hbar N_{i},$ while both transmittances entering the
vertex are proportional to $1/\tau _{\phi },$ the whole vertex is
proportional to the dephasing rate. The arrows make explicit that
transmittances are the product of a retarded (electron) and an advanced
(hole) Green's function. Obviously, one can sum the terms of Eq. (\ref
{eq-T-series}) to obtain the result of Eq.(\ref{eq-incoherent-transmittance}%
) for phenomenological decoherence.

Many of the results contained in the D'Amato-Pastawski paper for ordered and
disordered systems were extended in great detail in a series of papers by S.
Datta and collaborators and are presented in a didactic layout in a book 
\cite{cit-Datta}. In the next section, we will illustrate how the previous
ideas work by considering again our reference toy model for resonant
tunneling.

\subsection{Effects of decoherence in resonant tunneling}

After an appropriate decimation \cite{cit-Levstein-decim} the ``sample'' is
represented by a single state \cite{cit-GLBE2,cit-Jauho-book}. If we choose
to absorb the energy shifts into the site energies $\overline{E}%
_{0}=E_{0}+\Delta $, the Green's function is trivial

\begin{equation}
G_{0,0}^{R}=\frac{1}{\varepsilon -\overline{E}_{0}+\mathrm{i}(^{\mathrm{L}%
}\Gamma +\,\,^{\mathrm{R}}\Gamma +\,\,^{\phi }\Gamma )}.
\end{equation}
By taking the $\Gamma $'s independent on $\varepsilon $ in the range of
interest we get the ``broad-band'' limit. From now on we drop unneeded
indices and arguments. From this Green's function all the transmission
coefficients can be evaluated at the Fermi energy. 
\begin{align}
T_{\mathrm{R},\mathrm{L}}& =4\,^{\mathrm{R}}\Gamma \,\left| G_{0,0}\right|
^{2}\,^{\mathrm{L}}\Gamma , \\
T_{\phi ,\mathrm{L}}& =4\,^{\phi }\Gamma \,\left| G_{0,0}\right| ^{2}\ ^{%
\mathrm{L}}\Gamma ,\,\mathrm{and}  \notag \\
T_{\mathrm{R},\phi }& =4\,^{\mathrm{R}}\Gamma \,\left| G_{0,0}\right|
^{2}\,^{\phi }\Gamma .  \notag
\end{align}
We obtain the energy dependent total transmittance: 
\begin{equation}
T_{\mathrm{R},\mathrm{L}}(\varepsilon )=4\,\,^{\mathrm{R}}\Gamma \frac{1}{%
\left( \varepsilon -E_{0}\right) ^{2}+(^{\mathrm{L}}\Gamma +\,\,^{\mathrm{R}%
}\Gamma +\,\,^{\phi }\Gamma )^{2}}\,^{\mathrm{L}}\Gamma \left\{ 1+\frac{%
^{\phi }\Gamma }{^{\mathrm{L}}\Gamma +^{\mathrm{R}}\Gamma }\right\} .
\label{eq-Treson+DPmodel}
\end{equation}
The first term in the curly bracket is the coherent contribution while the
second is the incoherent one. We notice that the effect of the decoherence
processes is to lower the value of the resonance from its original one in a
factor:

\begin{equation}
\frac{\left( ^{\mathrm{L}}\Gamma+^{\mathrm{R}}\Gamma\right) }{(^{\mathrm{L}%
}\Gamma+\,\,^{\mathrm{R}}\Gamma+\,\,^{\phi}\Gamma)}
\end{equation}
In compensation, transmission at the resonance tails becomes increased.

It is interesting to note the results in the non-linear regime when the
voltage drop $e\mathsf{V}$ is greater than the resonance width $(^{\mathrm{L}%
}\Gamma+\,\,^{\mathrm{R}}\Gamma+\,\,^{\phi}\Gamma).$ If the new resonant
level lies between $\mu_{o}+$ $e\mathsf{V}$ and $\mu_{o},$ we can easily
compute the \textit{non-linear current }using $T_{\mathrm{R},\mathrm{L}%
}(\varepsilon ,e\mathsf{V})$. Notably, one gets that the total current does
not change as compared with that in absence of decoherent processes, i.e.:

\begin{align}
\mathsf{I}\frac{h}{2e} & =\int_{\mu_{o}}^{\mu_{o}+e\mathsf{V}}\widetilde
{T}_{\mathrm{R,L}}(\varepsilon,e\mathsf{V})\mathrm{d}\varepsilon=\int_{\mu
_{o}}^{\mu_{o}+e\mathsf{V}}T_{\mathrm{R,L}}^{o}(\varepsilon,e\mathsf{V})%
\mathrm{d}\varepsilon \\
& =4\pi\,^{\mathrm{R}}\Gamma\frac{1}{(^{\mathrm{L}}\Gamma+\,\,^{\mathrm{R}%
}\Gamma)}\,^{\mathrm{L}}\Gamma.
\end{align}
Thus, in this extreme quantum regime, the \textit{decoherence processes do
not affect the overall current}. Such relative ``stability'' against
decoherence is fundamental in the Integer Quantum Hall Effect \cite
{cit-IQ-Hall+dephasing} and should also be present for tunneling through
molecular states as both of them have a discrete spectrum. Notice that in
this last case the dependence of the escape rates on $e\mathsf{V}$ is
generally weak. Hence, the experimental value of the current allows an
estimation of the escape rate.

We learn important lessons from the case of resonant tunneling with the
inclusion of external degrees of freedom as decoherence:

1) The integrated intensity of the elastic (coherent) peak is decreased.

2) An inelastic current comes out to compensate this loss and maintains the
value of the total transmittance integrated over energy.

3) Decoherence \textit{broadens} both contributions to the resonance,
relaxing energy conservation.

The representation of the electron-phonon through complex self energies
damps the quantum interferences associated with repeated interactions with
some vibrational modes that would originate the polaronic states. In what
follows we will explore some simple Hamiltonian models where decoherence is
not introduced in such an early stage. Instead of calculating transition
rates, the approach will be to consider the many-body problem and to compute
the quantum amplitudes for each state in the Fock space.

\section{The electron-phonon models}

\subsection{A state conserving interaction model}

Let us consider a simple model that complements that of DP by providing an
explicit description of a single extended vibrational mode whose quanta in a
solid would be optical phonons,

\begin{equation}
\widehat{\mathcal{H}}_{ph}=\hbar\omega_{o}\widehat{b}_{{}}^{+}\widehat {b}.
\label{eq-H-phonon}
\end{equation}
The electrons are described by:

\begin{equation}
\widehat{\mathcal{H}}_{e}=\sum_{i=-\infty }^{\infty }E_{i}\widehat{c}_{i}^{+}%
\widehat{c}_{i}^{{}}-\sum_{j\left( \neq i\right) }^{{}}\left[ V_{i,j}%
\widehat{c}_{i}^{+}\widehat{c}_{j}^{{}}+V_{j,i}\widehat{c}_{j}^{+}\widehat{c}%
_{i}^{{}}\right] .  \label{eq-H-electron}
\end{equation}
The orbitals between $1$ and $N=L_{x}/a$ define our region of interest.
Orbitals at the edge are connected with the electrodes,\ where $V_{i,j}=V$,
through tunneling matrix elements $V_{0,1}=V_{1,0}\equiv V_{\mathrm{L}}$ and 
$V_{N,N+1}=V_{N+1,N}\equiv V_{\mathrm{R}}.$ The results will be simpler when 
$V_{\mathrm{L(R)}}\ll V$. Electrons and phonons are assumed to be coupled
through a local interaction:

\begin{equation}
\widehat{\mathcal{H}}_{e-ph}=\sum_{i=1}^{N}-V_{g}\widehat{c}_{i}^{+}\widehat{%
c}_{i}^{{}}(\widehat{b}_{{}}^{+}+\widehat{b}).  \label{eq-k-conservE-Ph}
\end{equation}
To build the electron-phonon Fock space we consider a single electron
propagating in the leads while the number $n$ of vibronic excitations is
well defined. While this election neglects the phonon mediated
electron-electron interaction, it still has non-trivial elements that are
the basis for the development of the concept of phonon laser (SASER) \cite
{cit-BrazJPhys}. This model will be called State Conserving Interaction
(SCI).

\begin{figure}
\begin{center}
\includegraphics*[width=10cm]{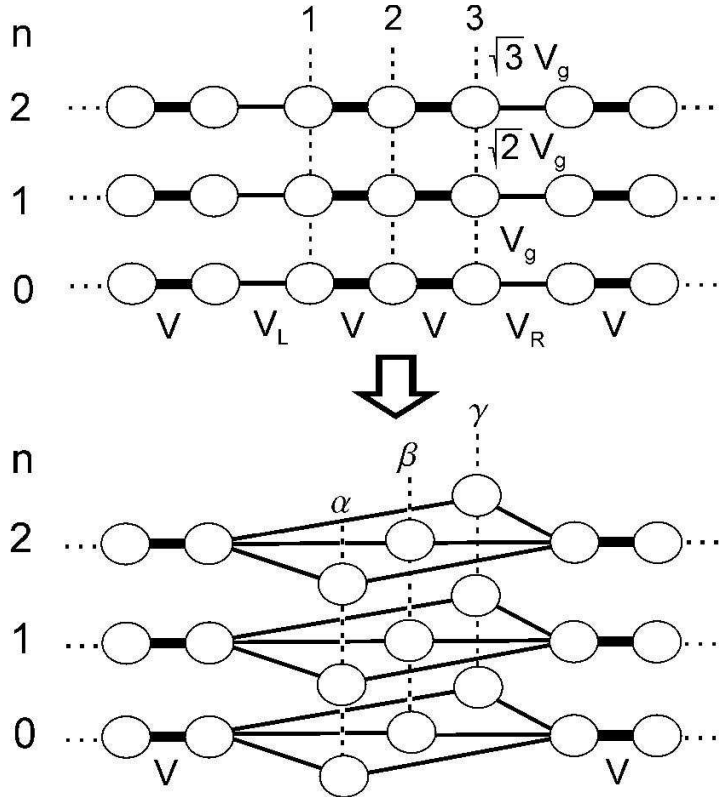}
\end{center}
\caption{Scheme for the problem of an electron plus a single phonon mode. The interacting system in the upper figure is transformed into polaronic modes associated to spatial eigenstates which interact weakly through the leads (lower figure).}
\label{fig-globalpolaron}
\end{figure}

The\textit{\ total energy}, which is conserved during the transport process,
is

\begin{equation*}
E=\varepsilon _{n}+n\hbar \omega _{o}\mathrm{\ \,\,with\,\,\,}0\leq
\varepsilon _{n}\leq \varepsilon _{F}.
\end{equation*}
Here $\varepsilon _{n}$ is the kinetic energy of the incoming electron on
the left lead and $n$ is the number of phonons before the scattering
process. A scheme of the complete electron-phonon Fock space is shown in
Fig. \ref{fig-globalpolaron}. It is clear that an electron that impinges
from the left when the well has $n$ phonons, can escape through the left or
right electrodes leaving behind a different number of phonons. Since these
are physically different situations, the outgoing channels with different
number of phonons are orthogonal and therefore cannot interfere. This is
represented in the fact that the sum of transmittances over final channels
satisfy unitarity.

The physical analysis of the excitations can be simplified resorting to a
new basis to refer the electron-phonon states. We notice that, if we neglect
the interaction with the leads, we can diagonalize the electronic
Hamiltonian without affecting the form of the electron-phonon interaction.
First we diagonalize the electronic system finding the annihilation
operators $\widehat{c}_{\alpha}^{{}}=\sum_{i=1}^{N}u_{\alpha,i}\widehat{c}%
_{i}^{{}}$ at eigenstates $\psi_{\alpha}$ with energies $E_{\alpha}$ and
wave functions $u_{\alpha,i}=\left\langle i\right| \left.
\alpha\right\rangle $

\begin{equation}
\widehat{\mathcal{H}}_{0}=\sum_{\alpha=1}^{N}E_{\alpha}\widehat{c}_{\alpha
}^{+}\widehat{c}_{\alpha}^{{}}\,\,\,\mathrm{and}\,\,\,\widehat{\mathcal{H}}%
_{e-ph}=\sum_{i=0}^{N}-V_{g}\widehat{c}_{\alpha}^{+}\widehat{c}%
_{\alpha}^{{}}(\widehat{b}_{{}}^{+}+\widehat{b})
\end{equation}

The Hamiltonian $\widehat{\mathcal{H}}_{e-ph}$ is just a linear field for
the harmonic oscillator where the field amplitude is proportional to the
density of the electrons that disturbs the lattice. Hence, the excitations,
called Holstein polarons, are easily obtained. The interesting point is that
with the proposed Hamiltonian one can diagonalize simultaneously every
electron subspace. i.e. in this model the phonons do not cause transitions
between electronic levels. The new excitations are described by 
\begin{equation}
\left( \widehat{a}_{k}^{+}\right) ^{n}\left| 0_{p,k}\right\rangle
=\sum_{n^{\prime }=0}^{\infty }\chi _{n,n^{\prime }}\left( \widehat{b}%
_{{}}^{+}\right) ^{n^{\prime }}\widehat{c}_{k}^{+}\left| 0\right\rangle
\end{equation}
which is valid at every electron space index $\alpha $. These operators
represent polarons in the energy basis analogous to the Holstein's local
polaron model \cite{cit-Wingreen}. The ``polaronic'' ground state is related
to the unperturbed state by $\left| 0_{p,k}\right\rangle =\sum_{n=0}^{\infty
}\chi _{0,n}\left( \widehat{b}_{{}}^{+}\right) ^{n}\widehat{c}_{k}^{+}\left|
0\right\rangle .\,$The polaronic energies are lower than those of electrons
plus phonons: 
\begin{equation}
E_{\alpha ,n_{{}}^{\prime }}=E_{\alpha }+\hbar \omega _{o}(n^{\prime }+%
\tfrac{1}{2})-\dfrac{\left| V_{g}\right| ^{2}}{\hbar \omega _{o}}\mathrm{\
with\,\,}n_{{}}^{\prime }=\left\langle \hat{a}^{+}\hat{a}\right\rangle .
\label{eq-polaron-energy}
\end{equation}
Since phonons do not produce transitions between electron energy states,
this model introduces decoherence through a \textit{State-Conserving
Interaction} (SCI). The lesson we learn from this is that by writing the
interactions in different basis we can choose the quantum numbers that are
conserved: e.g. local densities in the DP model and energy eigenstates in
the SCI model.

\subsection{Coherent and decoherent effects in electronic transport.}

The general problem of the electronic transport has been solved within the
formalism of Keldysh \cite{cit-GLBE2,cit-Jauho-book} in a FGR decoherent
approximation \cite{cit-DATTA-e-ph}. However, here we will solve the full
many-body problem. Figure \ref{fig-globalpolaron} makes evident that the
phonon emission/absorption can be viewed as a ``vertical'' hopping in a two
dimensional network. Once we notice that the Fock-space is equivalent to the
electron tight-binding model with an expanded dimensionality \cite
{cit-BrazJPhys}\cite{cit-Bonca}, we see that the transmittances can be
calculated exactly from the Schr\"{o}dinger equation. While excitations are
best discussed in the polaronic basis,\ for the electron transport it is
preferable the use of the asymptotic states.\ There, when the charge is
outside the interacting region, the electron-phonon product states
constitute the natural basis. Recently, this model has gained additional
interest as it has been used to explain inelastic effects in STM through
molecules \cite{cit-MingoPRL2000}, to study the transport in molecular wires 
\cite{cit-Kirczenow-e-ph} and to investigate Peierls's like distortions
induced by current in organic compounds \cite{cit-Ness-e-ph}.

A transport calculation is simpler if we prune the Fock space to include
only states within some range of $n$\ allowing a non-perturbative
calculation which can be considered variational in $n$. Thus, we are not
restricted to a weak e-ph coupling. To obtain the transmittances between
different channels several methods can be adopted. One possibility is to
solve for the wave function iteratively \cite{cit-Bonca}. An alternative is
to obtain Green's functions to get the transmittances. In this case, the
horizontal ``dangling chains'' in the Fock space can be eliminated through a
decimation procedure \cite{cit-D'Amato-Pastawski,cit-Levstein-decim}
introducing complex self-energies in the corresponding ``sites''.

\begin{figure}
\begin{center}
\includegraphics*[width=10cm]{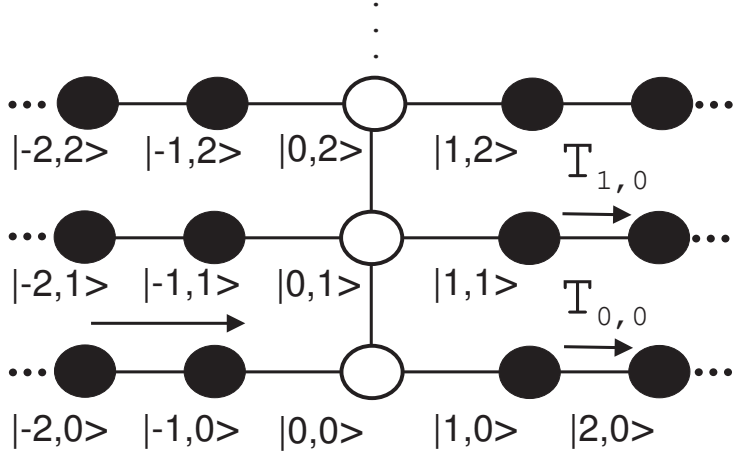}
\end{center}
\caption{Each site is a state in the Fock space: The lower row represents electronic states in different sites with no phonons in the well, the sites in black are in the leads. Higher rows correspond to higher number of phonons. Horizontal lines are hoppings and vertical lines are e-ph couplings. }
\label{fig-simpleModel}
\end{figure}

For simplicity, let us consider the case of a single state of energy $E_{0}$
in the region of interest which we will call the ``resonant'' state. It
could be interpreted as a HOMO or LUMO state depending on the situation. It
interacts with a dispersionless phonon mode, and\ it is coupled to a source
and drain of charge. The problem for an electron that tunnels through the
system can be mapped to the one-body problem shown in Fig. \ref
{fig-simpleModel}.

\begin{figure}
\begin{center}
\includegraphics*[width=12cm]{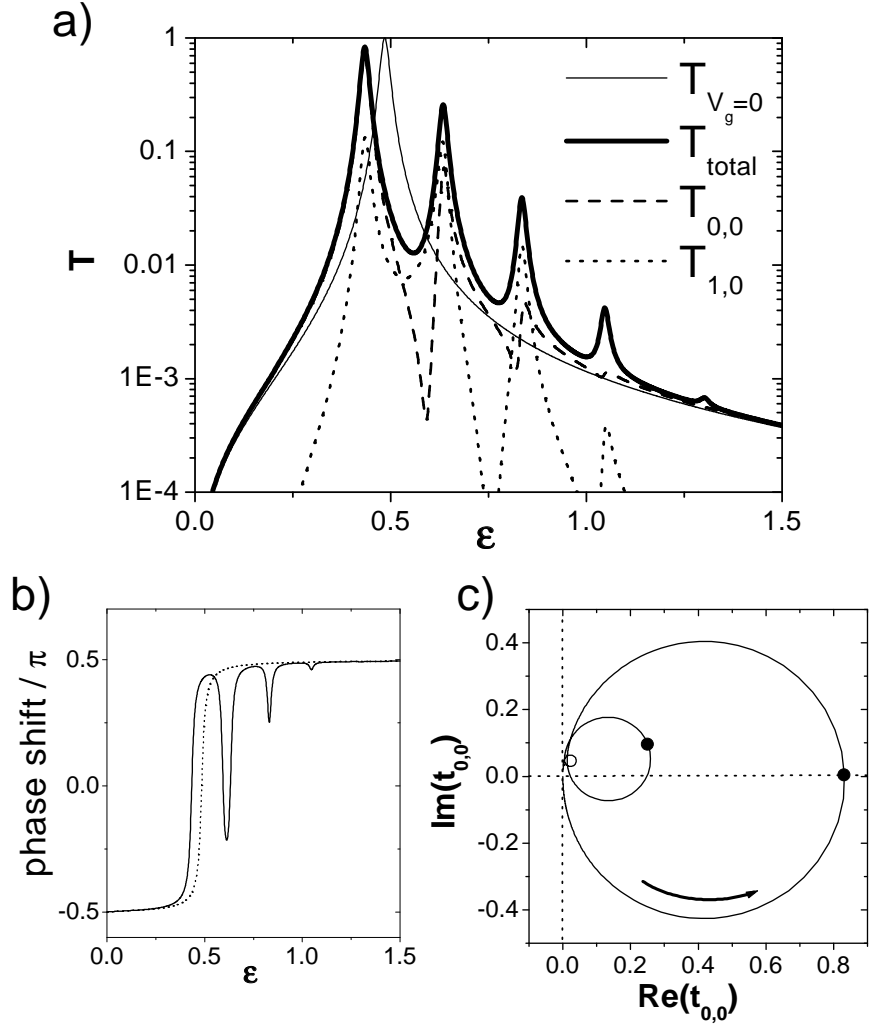}
\end{center}
\caption{a) Transmittance as a function of the incident
electronic kinetic energy. The total transmittance is shown with a thick solid line, the elastic component with a dashed line and $T_{1,0}$ with a dotted line. The non-interacting transmittance is also shown with a thin line. In b) the phase shift of the elastic transmittance as a function of
the energy and for the non-interacting case (dotted line) are shown. Figure c) shows the path of the elastic transmission amplitude in the complex plane when the electronic energy is changed. The full circles correspond to the first and second peaks shown in a). These results are obtained with $%
E_{0}=-1.5,$ $V_{\mathrm{L}}=V_{\mathrm{R}}=0.1,$ $V_{g}=0.1$ and $\hbar%
\protect\omega_{0}=$ $0.2.$ }
\label{fig-panelAntiRes}
\end{figure}

Let us consider an asymptotic incoming scattering state consisting of a wave
packet built with electron-phonon states in the left branch corresponding to 
$n_{0}$ phonons, i.e. an electron coming from the left while there are $%
n_{0} $ phonons in the well. When it arrives at the resonant site where it
couples to the phonon field, it can either keep the available energy $E$ as
kinetic energy or change it by emitting or absorbing $n$ phonons. Each of
these processes contributes to the total transmittance which is given by: 
\begin{equation}
T_{\mathrm{tot}}=\sum_{n}T_{n_{0}+n,n_{0}}^{{}}.
\end{equation}
In Fig. \ref{fig-panelAntiRes}-a we show the total transmittance (thick
line) for a case where $n_{0}$ is zero and the e-ph coupling is weak. There,
the appearance of satellite peaks at energies $E_{0}+n\hbar \omega _{0}$ can
be appreciated. To discriminate the processes contributing to the current,
we include with a dashed line, the \textit{elastic} contribution to the
transmittance, i.e. that due to electrons which are escaping to the right
without leaving vibrational excitations behind. This makes almost all of the
main peak and just decreasing portions of the satellite peaks. The elastic
contribution at the satellite peaks corresponds to a sum of \textit{virtual
processes} consisting of the emission of phonons followed by their immediate
reabsorption. The \textit{inelastic} component associated with the emission
of one phonon is shown with a dotted line.

While the total transmittance (thick line) shows a smooth behavior, the
elastic component (dashed line) exhibits a strong dip in the region between
the first two resonances. Therefore, almost all of the transmitted electrons
within this energy range will be scattered to the inelastic channels. This
sharp drop in the elastic transmittance is produced by a destructive
interference between the different possible ``paths'' in the Fock-space
connecting the initial and the final channels. These paths can be classified
essentially as a direct ``path'' between the incoming and outgoing channels
and the same path dressed with virtual emission and absorption processes.
The first inelastic component (dotted line) also shows a similar behavior in
the valley between the second and third resonances. The main factors that
control the magnitude of these \textit{antiresonances} are the escapes to
the leads and the \textit{e-ph} coupling. This \ concept\textit{\ } was
introduced in Refs.\cite{cit-DAmato AB,cit-Levstein-decim} to extend the
Fano-resonances \cite{cit-Fano} observed in spectroscopy to the problem of
conductance. Perfect antiresonances correspond to situations where both the
real and the imaginary part of the transmission amplitude are zero. In the
present case, such a perfect interference condition does not occur, and
there is a non-zero minimum transmission at the dips. Through a similar
analysis one could tailor the geometrical parameters of the system (as done
in Ref. \cite{cit-PRBFock2001}) to optimize the phonon emission.

An alternative way to appreciate this interference effect is to plot the
path followed by the transmission amplitude in the complex plane when the
kinetic energy of the incoming electron is changed \cite
{cit-HLee-phase-lapse,cit-Buttiker-Friedel-Sum-Rule}. A plot for the elastic
transmission amplitude is shown in Fig. \ref{fig-panelAntiRes}-c). For $%
\varepsilon =0$, there is no transmission. If the energy is increased, one
starts to follow the path in the figure anti-clockwise. After reaching the
point corresponding to the first resonance, the transmission starts to
decrease and the curve develops a turning near the origin. The first
antiresonance takes place at the point of minimum distance from the origin.

\begin{figure}
\begin{center}
\includegraphics*[width=10cm]{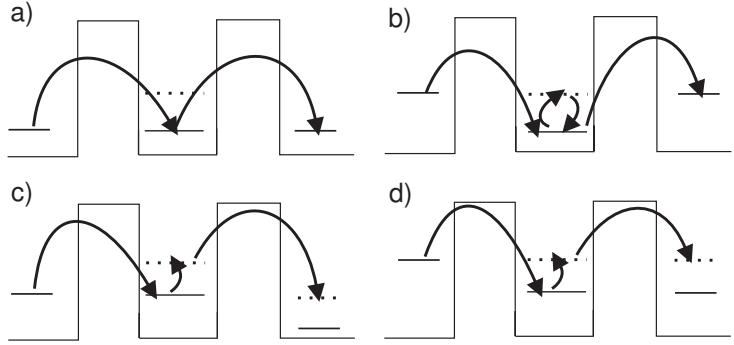}
\end{center}
\caption{Schematic representation of the processes that
lead to the elastic and inelastic components of the peaks in the total transmittance. Figures a) and b) correspond to the first and second peaks of the elastic transmittance respectively. Figures c) and d) correspond to the inelastic part. The well's ground state is represented by a solid line and the first excited polaron state by a dotted line. The final polaron states at the right of the well are also shown. These states have an energy equal to the incident electron kinetic energy. The final polaron state is depicted with a solid line for the elastic case and with a dotted line for the
inelastic situation. The level corresponding to the electron final energy is also shown as a solid line for this case.}
\label{fig-processes-pruned-Fock-space}
\end{figure}

In order to rationalize the main processes involved in the first two peaks,
let us represent them schematically in Fig.\ref
{fig-processes-pruned-Fock-space}. Panel (a) shows the standard elastic
process in which no phonon is emitted. Panel (b) is a notable effect that
occurs when the phonon emission is virtual. Notice that this virtual process
can have strong consequences. In Fig. \ref{fig-panelAntiRes} it\textbf{\ }%
produces an increase of the transmittance in almost \textit{two orders of
magnitude}. Panel (d) shows a real inelastic process which is expected to
give a peak if the initial energy satisfies $E=E_{o}+\hbar \omega _{o}$.
Notably, one must expect also a contribution when $\varepsilon =E_{o}$ which
corresponds to the virtual tunneling into the resonant state before emitting
a phonon as shown in panel (c).

Another interesting quantity that we can explore with the present formalism
is the phase of the\textbf{\ }transmitted electron through different
channels, 
\begin{equation}
\eta _{m,n}=\frac{1}{2\mathrm{i}}\ln \frac{G_{m,n}^{R}}{G_{n,m}^{A}},
\end{equation}
whose energy derivative gives information on the dynamics of the process.
Figure \ref{fig-panelAntiRes}-b shows the phase shift of the elastic
transmission probability. For reference, the phase shift in the absence of
e-ph interaction is also shown with a dotted line. It increases by $\pi $
over the width of the non-interacting transmission resonance. The same
occurs in the interacting case. However, it can be seen that each satellite
peak has associated a phase fluctuation. Instead of an increase in the phase
by $\pi $ which would occur for real resonant peaks, across each ``satellite
peak'' associated with the virtual processes, there is a phase dip that
results in consecutive resonances having phases close to -$\pi /2$. These
phase dips are a manifestation of the anti-resonances shown in Fig. (\ref
{fig-panelAntiRes}-a) for the elastic transmittance. For perfect zero
transmission points, one has an abrupt phase fall of $\pi $ instead of the
smooth phase slip shown in the Fig. (\ref{fig-panelAntiRes}-b) \cite
{cit-HLee-phase-lapse}.

Let us consider a more general case in which there are initially $n_{0}$
phonons in the scattering region. The vertical hopping matrix element
connecting states with $n_{0}$ and $n_{0}+1$ phonons is $\sqrt{n_{0}+1}V_{g}$%
. Then, under these conditions, the adimensional parameter, 
\begin{equation}
\widetilde{g}=\left( \frac{\left( \sqrt{n_{0}+1}\right) V_{g}}{\hbar \omega
_{o}}\right) ^{2}=\left( n_{0}+1\right) g,
\end{equation}
characterizes the strength of the e-ph interaction. It presents two regimes
according to the importance of this interaction.

\begin{figure}
\begin{center}
\includegraphics*[width=10cm]{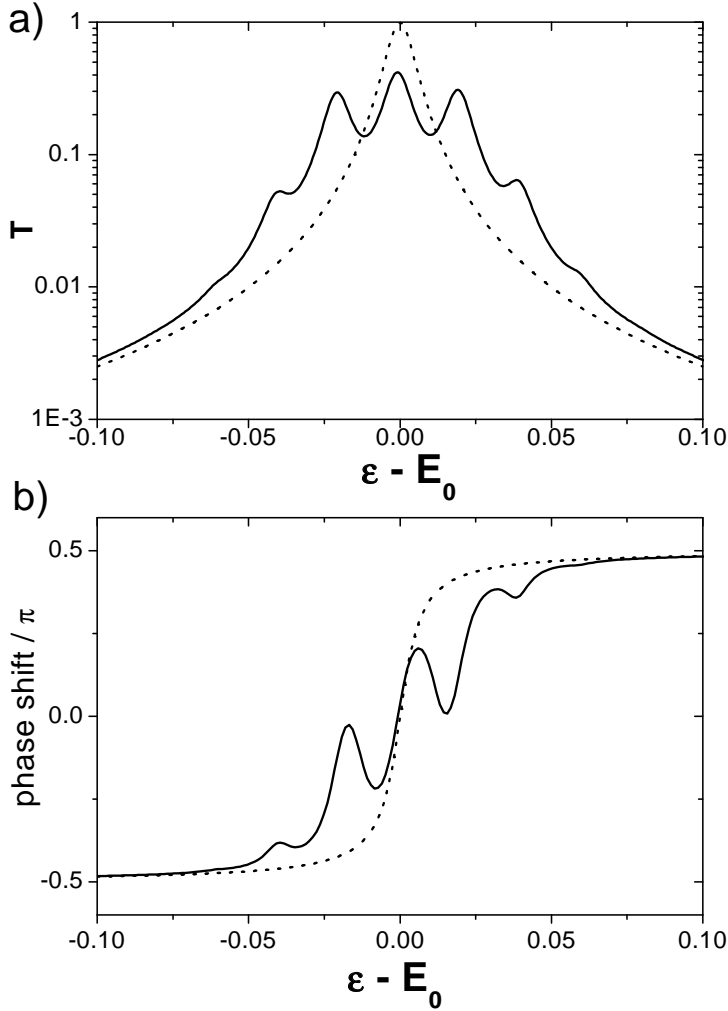}
\end{center}
\caption{a) Total transmittance as a function of the incident electronic kinetic energy. The solid line corresponds to the case in which there are 10 phonons in the well before the scattering process. The dotted line corresponds to the case in which there is no interaction with the vibrational degrees of freedom. b) Phase shift in units of $\protect\pi$
as a function of the incident electronic kinetic energy. The solid line is the phase shift for the elastic transmission. The dotted line corresponds to the non-interacting case. The parameters of the Hamiltonian in units of the hopping $V$ are: $E_{0}=0,$ $V_{L}=V_{R}=0.05,$ $V_{g}=0.004$ and $\hbar 
\protect\omega_{0}=0.02$.}
\label{fig-WeakCoupling}
\end{figure}

In the limit $\widetilde{g}\ll 1$ the ``vertical'' processes are in a
perturbative regime. The ``vertical hopping'' $\sqrt{n_{0}+1}V_{g}$ will not
be able to delocalize the initial state along this ``vertical direction''
and therefore the elastic contribution is the most important. In Fig. (\ref
{fig-WeakCoupling}-a), we show with a thick line the transmission
probability for a case where $g=0.25$ in presence of 10 phonons. Notice that
``satellite'' resonances corresponding to phonon emission and absorption are
separated by $\hbar \omega _{o}$ from the main resonance. For comparison,
the curve in absence of electron-phonon interaction is also included with a
dotted line. We see that, as with the DP model, the main resonance peak
decreases and presents a general broadening where we can recognize details
of the excitation structure of the phonon field. The phase shift as a
function of the electronic energy is shown in b). The solid line is the
phase shift for the elastic component of the total transmittance. There, we
can appreciate the phase fluctuations that appear even in the elastic
transmission.

On the other hand, if $\widetilde{g}\gtrsim1,$ the e-ph interaction is in
the non-perturbative regime. This can be achieved either by a strong $V_{g}$%
, or by a high $n$. The second situation would correspond to the case of
high temperatures or a far from equilibrium phonon population. In that case,
the total transmittance can be obtained by summing the transmittances for
the different possible initial conditions weighted by the appropriate
thermal factor \cite{cit-Haule-Bonca-1999}. In any of these situations, the
resulting strong \textit{e-ph} coupling leads to a breakdown of \
perturbation theory. A similar case is found for the quantum dots studied in
Ref. \cite{cit-Bastard}. There, in contrast to the situation in bulk
material, the studied quantum dots show a strong \textit{e-ph} coupling. In
molecular wires, where elementary excitations (e or ph) are highly confined,
one can expect a similar scenario.

\begin{figure}
\begin{center}
\includegraphics*[width=10cm]{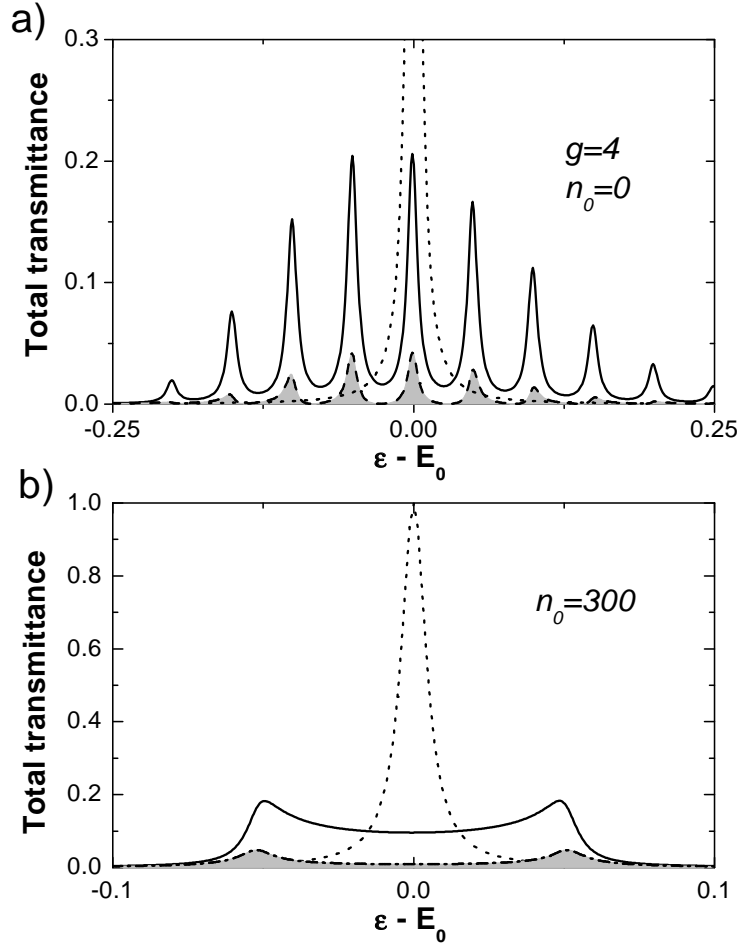}
\end{center}
\caption{a) and b) show the total transmittance as a
function of the incident kinetic electronic energy. The area under the elastic transmittance is shaded in gray. As before, the dotted line corresponds to the non-interacting case. In a), the solid line corresponds to the case in which there are no phonons in the well before the
scattering process and $V_{g}=0.100$, $\hbar\protect\omega_{0}=0.05$. In b)
, the solid line corresponds to the case in which there are 300 phonons in the well before the scattering process and $V_{g}=0.0015$ and $\hbar\protect%
\omega _{0}=5.010^{-4}$. The parameters in units of the hopping $V$ are: $E_{0}=0$, $V_{L}=V_{R}=0.05.$}
\label{fig-StrongCoupling}
\end{figure}

In Fig. (\ref{fig-StrongCoupling}-a), we show the transmittance as a
function of the incident electron kinetic energy for the case where $V_{g}$
is strong ($\widetilde{g}=4$ ) and there are no phonons in the well before
the scattering process. In contrast to the perturbative regime, where the
energy shift of the peaks ($-V_{g}^{2}/\hbar \omega _{o}=-g\hbar \omega _{o}$%
) is small, here the energy shift is important and the inelastic
contributions to the total transmittance dominate. In the limit of weak
coupling between the scattering region and the leads, the transmission
probability through the channel with $n$ phonons is: $\exp (-g)g^{n}/n!$.

Figure (\ref{fig-StrongCoupling}-b) shows the total transmittance as a
function of the incident electron kinetic energy for an extreme case where
there are 300\textbf{\ }phonons in the well before the scattering process.%
\textbf{\ }The electron can absorb or emit as many phonons, $N_{eff}=4\sqrt{n%
}V_{g}/\hbar \omega _{o},$ as allowed by the interaction strength. This
means that the phonon spectrum, weighted on the local electron-phonon state $%
n$, will be quite independent on the details of the spectral densities at
the far end of the effective ``vertical'' chain i.e. the states $n\pm
N_{eff} $. Hence, the spectral density will show the typical form of a one a
dimensional band even when we take $N_{eff}=4(V_{g}/\hbar \omega _{o})\sqrt{n%
}\ll N$. This feature manifests in the transmittance presented in Fig. \ref
{fig-StrongCoupling}-b). The unperturbed resonant peak is a dotted line
which contrasts with the total transmission in presence of phonons shown
with a continuous line. Its trace follows the structure of the phonon
excitation with the typical square root divergences at the band edges
smoothed out by the inhomogeneity in the hopping elements and the
uncertainty introduced by the escape to the leads. \smallskip

\section{The solution of Time Dependence}

Finally, we would like to present the basic features of time dependent
transport. This is relevant since a coherent ``time of flight'' should be
shorter than any conformational correlation time \cite{cit-Ratner-TIME}. The
basic idea to get the physics of time dependent phenomena is to obtain the
evolution of an arbitrary initial boundary condition of the form $\left[
\psi _{{}}(X_{j})\psi _{{}}^{\ast }(X_{k})\right] _{\mathrm{source}}.$ Since
here $X_{j}=(\mathbf{r}_{j},t_{j})$ with general $t_{j},$ this essentially
generalizes a density matrix which introduces temporal correlations that
define the energy of this injected particle. The ``density'' at a later is
obtained from the exact solution of the Schr\"{o}dinger equation:

\begin{equation}
\left[ \psi (X_{2})\psi ^{\ast }(X_{1})\right] =\hbar ^{2}\int \int
G^{R}(X_{2,}X_{j})\left[ \psi (X_{j})\psi ^{\ast }(X_{k})\right] _{\mathrm{%
source}}G^{A}(X_{k,}X_{1})\mathrm{d}X_{j}\mathrm{d}X_{k}
\label{eq-density-evolution}
\end{equation}
In order to establish a correspondence with the Danielewicz solution to the
Schr\"{o}dinger\ equation \ in the Keldysh formalism, we used the continuous
variable Green's function which is related to the exact discrete one in the 
\textit{open} system by 
\begin{equation}
G^{R}(\mathbf{r}_{i,}\mathbf{r}_{j},\varepsilon
)=\sum_{k,l}G_{k,l}^{R}(\varepsilon )\varphi _{k}(\mathbf{r}_{i})\varphi
_{l}^{\ast }(\mathbf{r}_{k}).  \label{G-continua-discreta}
\end{equation}
Now, the key is to recognize that in any Green's function, a macroscopically
observable time is $t=\frac{1}{2}\left[ t_{j}+t_{k}\right] $ (time center).
Its Fourier transform is an observable frequency $\omega $. Meanwhile,
energies are associated with internal time differences $t_{j}-t_{k}$ $\ $%
(time chords). Within this scheme, the time correlated initial condition
(source), determining the occupation of a local orbital is expressed in
terms of the time independent local density of states $N_{i}(\varepsilon ):$ 
\begin{align}
\left[ u_{i}(t)u_{i}^{\ast }(t)\right] _{\mathrm{source}}& =\int \frac{%
\mathrm{d}\varepsilon }{2\pi \hbar }\int \mathrm{d}\delta t\,\,\left[
u_{i}^{{}}(t+\delta t/2)u_{i}^{\ast }(t-\delta t/2)\right] _{\mathrm{source}%
}\exp [\mathrm{i}\varepsilon \delta t]  \notag \\
& =\int \mathrm{d}\varepsilon N_{i}(\varepsilon )\mathrm{f}(\varepsilon ,t),
\label{eq-density-occupation}
\end{align}
and the occupation factor $\mathrm{f}(\varepsilon ,t).$ Introducing this
notation in Eq.(\ref{eq-density-evolution}) one can identify the dynamical
transmittances $T_{i,j}(\varepsilon ,\omega ).$ We refer to \cite{cit-GLBE2}
for a detailed manipulation of the time integrals. The basic result is

\begin{equation}
T_{i,j}(\varepsilon,\omega)=2\Gamma_{i}(\varepsilon)G_{i,j}^{R}(\varepsilon +%
\tfrac{1}{2}\hbar\omega)2\Gamma_{j}(\varepsilon)G_{j,i}^{A}(\varepsilon -%
\tfrac{1}{2}\hbar\omega).  \label{eq-T(w)}
\end{equation}
The time dependent transmittance is then:

\begin{equation}
T_{i,j}(\varepsilon ,t)=\int T_{i,j}(\varepsilon ,\omega )\exp [-i\omega t]%
\frac{\mathrm{d}\omega }{2\pi },  \label{eq-T(t)}
\end{equation}
where one recovers our old steady state transmittance as: 
\begin{equation}
T_{i,j}(\varepsilon )\equiv T_{i,j}(\varepsilon ,\omega =0)=\int_{-\infty
}^{t}T_{i,j}(\varepsilon ,t-t_{i})\mathrm{d}t_{i}=\int_{t}^{\infty
}T_{i,j}(\varepsilon ,t_{f}-t)\mathrm{d}t_{f}.  \label{eq-Tsteady=Ttimedep}
\end{equation}
Equation (\ref{eq-density-evolution}) then becomes 
\begin{equation}
\mathsf{I}_{j}(t)=\frac{2e}{h}\int \mathrm{d}\varepsilon \sum_{i}\left[
T_{i,j}(\varepsilon )\mathrm{f}_{j}(\varepsilon ,t)-\int_{-\infty }^{t}%
\mathrm{d}t_{i}T_{j,i}(\varepsilon ,t-t_{i})\mathrm{f}_{i}(\varepsilon
,t_{i})\right] ,
\end{equation}
which is the Generalized Landauer-B\"{u}ttiker Equation (GLBE) \cite
{cit-GLBE1,cit-GLBE2}. According to Eq.(\ref{eq-Tsteady=Ttimedep}), the
first term accounts for the particles that are leaving the reservoir at site 
$j$ at time $t$ to suffer a dephasing collision at some future time at
orbital $i.$ The second accounts for particles that having had a previous
dephasing collision at time $t_{i}$ at the $i$ orbital reach site $j$ at
time $t$.

In the GLBE formulation the essential features of time dependence in
transport is contained in the transmittances. Since the spectrum is
continuous, we can keep the lowest order in the frequency expansion e.g. 
\begin{equation}
T_{i,j}(\varepsilon ,\omega )\simeq \dfrac{T_{i,j}(\varepsilon )}{1-\mathrm{i%
}\omega \tau _{1}+(\omega \tau _{2}^{{}})^{2}+...}.  \label{eq-T(w) approx}
\end{equation}
According to Eq. (4.5) in Ref. \cite{cit-GLBE2} the propagation time $\tau
_{P}$ is identified with the first significant coefficient in this
expansion. Typically, it results the first order, 
\begin{align}
\tau _{P}& =\frac{\mathrm{i}\hbar }{2}\left[ G_{i,j}^{R}(\varepsilon )\dfrac{%
\partial }{\partial \varepsilon }G_{i,j}^{R}(\varepsilon
)^{-1}+G_{j,i}^{A}(\varepsilon )\dfrac{\partial }{\partial \varepsilon }%
G_{j,i}^{A}(\varepsilon )^{-1}\right]  \label{eq-tau_P} \\
& =-\frac{\mathrm{i}\hbar }{2}\dfrac{\partial }{\partial \varepsilon }\ln %
\left[ \dfrac{G_{i,j}^{R}(\varepsilon )}{G_{j,i}^{A}(\varepsilon )}\right] 
\notag
\end{align}
which can be identified with the Wigner time delay. This propagation time
was evaluated in various simple systems in Ref. \cite{cit-GLBE2} recovering
the ballistic and diffusive times for clean and impure metals respectively.
In a double barrier system, in the \textit{resonant tunneling} regime, the
propagation time is determined by the life-time inside the well. In fact,
using the functions of Subsection 2.3, one gets for the propagation through
the resonant state: 
\begin{equation}
\tau _{P}=\dfrac{\hbar }{2(^{L}\Gamma +^{R}\Gamma )}.
\end{equation}
From these considerations, we see that $\tau _{P}$ \ represents a limit to
the response in frequency (admittance) of the device. Typically, one gets $%
\mathsf{G}_{\omega }=\mathsf{G}_{0}/(1-\mathrm{i}\omega \tau _{P})$. This is
in fair agreement with the experimental results \cite{cit-time/exp:Sollner}.

As an striking example, we mention the ``simple'' case of \textit{tunneling
through a barrier }\cite{cit-Landauer-time} of length $L$ and height $U$
exceeding the kinetic energy $\varepsilon $ of the particle. For barriers
long enough the expantion is dominated by the second order term in Eq. (\ref
{eq-T(w) approx}) and one gets: 
\begin{equation}
\tau _{P}=L/\sqrt{\frac{2}{m}(U-\varepsilon )},
\end{equation}
which, within our non-relativistic description, can be extremely short
provided that the barrier is high enough. This is the time that one has to
compare with vibronic and configurational frequencies \cite
{cit-Ratner-tuneling time}.

The general propagation times can also be calculated in more complex
situations such as disordered systems \cite{cit-Prigodin} and those affected
by incoherent interactions \cite{cit-GLBE2}.

\smallskip

\section{Perspectives}

We have presented the general features of quantum transport in mesoscopic
systems. There, interactions with environmental degrees of freedom introduce
complex phenomena whose overall effect is to decrease the clean
interferences expected for an isolated sample. Such degradation can be
accounted for as decoherence. We introduced various simple models which
present decoherence. We started presenting the simplest phenomenological
models of B\"{u}ttiker, by discussing its connection with the Hamiltonian
description of the DP model. Various simple models for resonant tunneling
devices including electron-phonon interactions were introduced. We feel that
they contain the essential coherent and decoherent effects induced by the
vibronic degrees of freedom.

Reference \cite{cit-D'Amato-Pastawski} showed that in disordered systems,
the effective transmittance away from the resonances keeps its form as a
superposition of tails from different resonances. This applies to both the
DP and SCI models. This justifies the simplification of using a single
resonant state.

Our results for this model make quite clear how complex many body
interactions result in the loss of the simple interferences of one body
description. Essentially, each external degree of freedom coupled with the
electronic states leads to two situations producing decoherence:

1) By \textit{real} emission or absorption of phonons, they open an
additional scattering channels contributing incoherently to the transport.
This situation is represented in the DP decoherence model, as well as in the
various polaronic models presented here. Those are real processes which can
be detected by measuring a change in the bath state (phonon models) or
energy dissipated (DP model) within the sample.

2) However, even when the electron-phonon processes might be \textit{virtual}
, they would add new alternatives to the quantum phase of the outgoing
state. This strongly limits an attempt to control the electron phase and
hence is manifested as decoherence.

The study of the coherent component in the presence of decoherent processes
gives a first, though imperfect, hint to the conditions of the transport
processes. In particular, it can show how decoherence can affect the
interference between different propagation pathways. Once they are summed
up, the conductance, which is a square modulus, would manifest the effect of
a diminished interference. This approach was adopted in a recent work \cite
{cit-Medina-Pastawski} that addressed the delocalizing effect of decoherence
on transport in the variable range hopping regime.

It is also worthwhile to mention that the randomization of the quantum phase
introduced by the virtual processes can be as much effective \cite
{cit-Jalabert-Pastawski} as those involving a real energy exchange in
producing decoherence. In fact, it has been found recently that this
practical uncertainty is the mechanism through which quantum chaos
contributes to dissipation and irreversibility \cite{cit-cook-LORENTZ}. Such
quantum chaotic systems have intrinsic decoherence time scales which
contrast with the extreme quantum regime of a resonant tunneling where
decoherence, if any, is controlled by the environment.

We want to close this section mentioning the connections of the results in
Section 4 with other ongoing research on various hot issues of electronic
transport. While the problem of tunneling times \cite
{cit-time-chains,cit-Landauer-time} is a controversial one, it has important
practical aspects \cite{cit-time/exp:Sollner}. Indeed, molecular electronics
opens the whole issue of quantum dynamical processes to a fresh
consideration. One aspect is the effect of decoherence on frequency
response. A related issue under study is the interconnection between
decoherence times, irreversibility and dynamical chaos \cite
{cit-Jalabert-Pastawski}. Having shown the subtle relation between spectral
properties and time dependences, one foresees that our models can provide
new insight to this topic. The analysis of the consequences of the spectral
complexity of many-body systems is a fully unexplored field ahead. Once
again, technology pushes us to the frontiers of the conceptual understanding
of Quantum Mechanics and Statistical Physics.

\section{Acknowledgments}

We acknowledge P. R. Levstein and F. Toscano for critical discussions. We
acknowledge financial support of bi-national programs of Antorchas-Vitae and
CONICET-CONICIT as well as local support from SeCyT-UNC. HMP and LEFFT are
affiliated with CONICET.

\end{document}